\documentstyle[12pt]{article}

\begin{document}

\textheight 9.0in
\topmargin -0.5in
\textwidth 6.5in
\oddsidemargin -0.01in
\def\singlespace {\smallskipamount=3.75pt plus1pt minus1pt
                  \medskipamount=7.5pt plus2pt minus2pt
                  \bigskipamount=15pt plus4pt minus4pt
                  \normalbaselineskip=15pt plus0pt minus0pt
                  \normallineskip=1pt
                  \normallineskiplimit=0pt
                  \jot=3.75pt
                  {\def\smallskip {\vskip\smallskipamount}}
                  {\def\medskip   {\vskip\medskipamount}}
                  {\def\bigskip   {\vskip\bigskipamount}}
                  {\setbox\strutbox=\hbox{\vrule
                    height10.5pt depth4.5pt width 0pt}}
                  \parskip 7.5pt
                  \normalbaselines}
\def\middlespace {\smallskipamount=5.625pt plus1.5pt minus1.5pt
                  \medskipamount=11.25pt plus3pt minus3pt
                  \bigskipamount=22.5pt plus6pt minus6pt
                  \normalbaselineskip=22.5pt plus0pt minus0pt
                  \normallineskip=1pt
                  \normallineskiplimit=0pt
                  \jot=5.625pt
                  {\def\smallskip {\vskip\smallskipamount}}
                  {\def\medskip   {\vskip\medskipamount}}
                  {\def\bigskip   {\vskip\bigskipamount}}
                  {\setbox\strutbox=\hbox{\vrule
                    height15.75pt depth6.75pt width 0pt}}
                  \parskip 11.25pt
                  \normalbaselines}
\def\doublespace {\smallskipamount=7.5pt plus2pt minus2pt
                  \medskipamount=15pt plus4pt minus4pt
                  \bigskipamount=30pt plus8pt minus8pt
                  \normalbaselineskip=30pt plus0pt minus0pt
                  \normallineskip=2pt
                  \normallineskiplimit=0pt
                  \jot=7.5pt
                  {\def\smallskip {\vskip\smallskipamount}}
                  {\def\medskip   {\vskip\medskipamount}}
                  {\def\bigskip   {\vskip\bigskipamount}}
                  {\setbox\strutbox=\hbox{\vrule
                    height21.0pt depth9.0pt width 0pt}}
                  \parskip 15.0pt
                  \normalbaselines}

%\rightline{OSU-PHYS-178}
%\rightline{WU-AP/115/00}
%\rightline{YITP-00-56}
\bigskip
\begin{center}
{\bf {\Large Naked Singularities and Quantum Gravity}}

\smallskip

%{\bf { - Interpreting the Quantum Divergence in Spherical
%Collapse - }}

\bigskip\

{\bf Tomohiro Harada$^{a}$\footnote{e-mail address:
harada@gravity.phys.waseda.ac.jp},
Hideo Iguchi$^{b}$\footnote{e-mail address: iguchi@th.phys.titech.ac.jp},
Ken-ichi Nakao$^{c}$\footnote{e-mail address:
knakao@sci.osaka-cu.ac.jp},
\break
T. P. Singh$^{d,e}$\footnote{e-mail address: tpsingh@mailhost.tifr.res.in},
Takahiro Tanaka$^{d}$\footnote{e-mail address:
tanaka@yukawa.kyoto-u.ac.jp}
and Cenalo Vaz$^{f}$\footnote{e-mail address: cvaz@ualg.pt}}

\bigskip

{\it $^{a}$Department of Physics, Waseda University}

{\it Ohkubo, Shinjuku, Tokyo 169-8555, Japan}

\medskip

{\it $^{b}$Department of Physics, Tokyo Institute of Technology}

{\it
 Oh-Okayama, Meguro-ku, Tokyo 152-8550, Japan}

\medskip

{\it $^{c}$Department of Physics, Osaka City University}

{\it Osaka 558-8585, Japan}

\medskip

{\it $^{d}$Yukawa Institute for Theoretical Physics}

{\it Kyoto University, Kyoto 606-8502, Japan}

\medskip

{\it $^{e}$Tata Institute of Fundamental Research}

{\it Homi Bhabha Road, Mumbai 400 005, India.}

\medskip\

{\it $^{f}$Unidade de Ciencias Exactas e Humanas}

{\it Universidade do Algarve, Faro, Portugal}

\medskip

\end{center}

\newpage
\begin{abstract}
There are known models of spherical gravitational collapse in which
the collapse ends in a naked shell-focusing singularity for some
initial data. If a massless scalar field is quantized on the classical
background provided by such a star, it is found that the outgoing
quantum flux of the scalar field diverges in the approach to the
Cauchy horizon. We argue
that the semiclassical approximation (i.e. quantum field theory on a classical
curved background) used in these analyses ceases to
be valid about one Planck time before the epoch of naked singularity
formation, because by then the curvature in the central region of the
star reaches Planck scale. It is shown that during the epoch in which the
semiclassical approximation is valid, the total emitted energy is about one
Planck unit, and is not divergent. We also argue that back reaction
in this model does
not become important so long as gravity can be treated classically.
It follows that the further
evolution of the star will be determined by quantum gravitational
effects, and without invoking quantum gravity it is not possible to
say whether the star radiates away on a short time scale or settles
down into a black hole state.
\end{abstract}

\newpage
\middlespace

\section{Introduction}

If a spherical star collapses to form a naked shell-focusing singularity, the
nature and magnitude of the energy emission accompanying such a collapse is of
interest. This emission may broadly be classified into three categories -
classical \cite{Dad} (where both matter and gravitational degrees of freedom are
treated classically), semiclassical (where only the matter degrees of
freedom are quantized on a classical curved background), and quantum gravitational
(where both matter and gravity degrees of freedom are quantized).

A priori, it is not possible to say which of these three kinds of emission will
dominate. In the present paper we bring together and analyze some of the
results concerning the semiclassical emission. It is known that the quantum Hawking
radiation is divergent as the Cauchy horizon is approached. We provide a fresh analysis
of the divergence of the quantum flux on the Cauchy horizon, and derive the
emission spectrum for a massless scalar field on the spherical dust background.
The divergence of the quantum flux has been observed in various models \cite{For},
\cite{His}, \cite{Bar}, \cite{Har} and the naked singularity spectrum was studied by
Vaz and Witten \cite{Vaz} and by Harada et. al. \cite{Har}.

The principal aim of the present paper is to show that during the epoch in which
the semi-classical analysis is valid, the total emitted energy is about one Planck unit
and is not divergent. If follows that, during this epoch, the back reaction is also
unimportant for an astrophysical star because the total radiated energy is small compared
with the mass of the star. Now the semiclassical approximation ceases to be valid about
one Planck time before the formation of a Cauchy horizon because by then the curvature
in the central region of the star reaches the Planck scale. Therefore further evolution
of the star, i.e., during the last Planck time before the formation of the Cauchy
horizon, will be determined by quantum gravity. An important consequence is that a
star collapsing to a naked singularity enters the quantum gravitational phase with most of
its mass intact. Because only about a Planck mass has been radiated away until then,
semi-classical techniques cannot indicate whether the star radiates away a significant portion
of its total mass on a short time scale or settles down into a black hole state.

\section{Radiated Power}

When a massless scalar field is quantized on the background of a collapsing spherical star,
the radiated power can be calculated in the geometric optics approximation. The expression
for the radiated power is \cite{For}
\begin{equation}
P =  {\hbar \over 24\pi}  \left[ {3\over 2} \left( {{\cal G}''\over
      {\cal G}'}\right)^{2} - {{\cal G}'''\over {\cal G}'}\right]=
{\hbar\over 48\pi}\left [ \left ( {{\cal G}''\over {\cal G}}\right)^{2}
- 2 \left({\cal G}''\over {\cal G }\right ) ' \right ],
\label{pow}
\end{equation}
{\it or}
\begin{equation}
\hat{P} =  {\hbar \over 48\pi}  \left( {{\cal G}''\over
      {\cal G}'}\right)^{2},
\label{hatpow}
\end{equation}
where $P$ and $\hat{P}$ refer to a minimally coupled scalar field, $\phi$, and a conformally
coupled scalar field, $\hat{\phi}$, respectively. The difference is due to the
nature of the coupling to the gravitational field and is seen to be only a total divergence
which does not contribute to the total radiated energy under some physically reasonable
assumptions. In this sense, it may be assumed that it is the radiated power for the
conformally coupled scalar field that is relevant for the observed energy flux because
this flux is positive definite. $P$ and ${\hat P}$ are the semiclassical contribution to the
expectation value of the off-diagonal component of the stress-energy tensor, $\langle
T_{UU}\rangle$. We note that it is entirely expressed in terms of the function ${\cal
G} (U)$, which maps ingoing null rays, $V=$constant, from ${\cal I}^{-}$, to
outgoing null rays $U=$constant on ${\cal I}^{+}$, {\it i.e.,} $V={\cal G}(U)$. This
description of the radiated energy in terms of null rays is a consequence of the fact that
the scalar field modes on the null infinites are most naturally expressed in terms of the
null coordinates, $U$ and $V$ there \cite{For}. The aforementioned divergence in the
radiated power is observed when the expressions (\ref{pow}) or (\ref{hatpow}) are evaluated
for the case of a spherical star collapsing to a naked singularity. No such divergence is
observed for a star collapsing to a black hole. In the following section, we display ${\cal G}
(U)$ for both cases. In the former the function is polynomial, in the latter it is exponential.
This key difference leads not only to the divergent flux but to a non-thermal spectrum as the
star collapses to form a naked singularity.

\section{The Dust Solution and the Map ${\cal G}(U)$}

The spherical collapse of dust, as described by the Tolman-Bondi model, leads to the formation of
a naked singularity for some initial data \cite{Chr}, \cite{New}, \cite{Wau} \cite{Dwi}, \cite{Sin}.
One can quantize a massless scalar field on the background of such a spacetime, and calculate the
quantum power radiated in the low angular momentum modes, using the geometrical optics approximation
described in the previous section.  This has been done for the self-similar dust model \cite{Bar},
for non-self-similar dust collapse \cite{bsv} and for a $C^{\infty}$ model \cite{Har}, all of
which are globally naked models. There is some indication from lower dimensional models that
the locally naked models do not radiate catastrophically \cite{vaz2}.

The marginally bound Tolman-Bondi metric  is given by
\begin{equation}
ds^{2}=dt^{2}-R'^{2}dr^{2}-R^{2}d\Omega^{2}.
\end{equation}
The evolution of the area radius is given by
\begin{equation}
R^{3/2}=r^{3/2}-{3\over 2}\sqrt{F(r)}t,
\end{equation}
where $F(r)$ is the mass function, and the density evolution is determined by the equation (we
set $G=c=1$)
\begin{equation}
8\pi\rho(t,r)={F'\over R^{2}R'}.
\end{equation}

The cloud is assumed to extend up to some comoving radius $r_{b}$ and is matched to a Schwarzschild
exterior. One can construct null coordinates  $(u,v)$ in the interior and the Eddington-Finkelstein
null coordinates $(U,V)$ in the exterior region.

The map ${\cal G}(U)$ defined in the previous section can be calculated, once an interior model for
the collapsing spherical dust star  is given. Let us first consider the self-similar model \cite{Bar}.
In this case, the mass function is given by $F(r)=\lambda r$ and we have the scaling $R=r$ at the
singular epoch, $t=0$. The density at the center evolves with time as
\begin{equation}
8\pi\rho={4\over 3t^{2}}
\label{den}
\end{equation}
becoming infinite at the singular epoch $t=0$.

At the boundary of the star, the Schwarzschild coordinates $T,R$ are related to the Tolman-Bondi
coordinates by
\begin{eqnarray}
T_b(t)~~ &=&~~  t~ -~ {2 \over {3\sqrt
{2M}}}r_b^{3/2}~ -~ 2 \sqrt{{2M R_b}}~ +~ 2M\ln |{{{\sqrt{R_b}} +
{\sqrt{2M}}} \over {{\sqrt{R_b}} - {\sqrt{2M}}}}|,\cr
 R_b(t)~~ &=&~~ r_b \left[
1 - a {t \over r_b} \right]^{2/3}.
\label{tee}
\end{eqnarray}
whence the Eddington-Finkelstein null coordinates become
\begin{eqnarray}
U_b(y)~~ &=&~~ -~ {{r_b} \over a} y^3~ -~  {4 \over 3} a r_b y~ -~ r_b y^2 ~ -~
{8\over 9} a^2 r_b \ln |3y/2a - 1|,\cr
V_b(y)~~ &=&~~ -~ {{r_b} \over a} y^3~ -~  {4 \over 3}
a r_b y~ +~ r_b y^2 ~ +~ {8\over 9} a^2 r_b \ln |3y/2a + 1|.
\label{ooo}
\end{eqnarray}
where $y=\sqrt{R/r}$ and $a=3\sqrt{\lambda}/2$.

Tracing a ray $V$ = constant from ${\cal I}^-$ through the cloud and out again toward
${\cal I}^+$, one finds that, for the self-similar model forming a naked singularity, the map
${\cal G}(U)$ is given by \cite{Bar}
\begin{equation}
{\cal G}(U)=  A  -  B  (U_0-U)^{1\over \gamma}
\label{nsm}
\end{equation}
where $\gamma$ is a positive constant, less than unity, which is determined by the only free
parameter, $\lambda$, in the self-similar model. Above, $U_{0}$ is the Cauchy horizon. On the
other hand, if the self-similar star collapses to form a black hole, the map is given by
\cite{sva}
\begin{equation}
{\cal G}(U)=  A  -  B  \exp\left[ -U/4M\right],
\label{bhm}
\end{equation}
where $M=F(r_{b})/2$ is the total mass of the star.

\section{Spectrum of Created Particles}

Let us evaluate the expectation value of the number operator of the created particles, in the
Minkowski in-vacuum. This is given by
\begin{equation}
\langle 0|N(\omega)|0\rangle=  \int_{0}^{\infty} d\omega ' ~|\beta(\omega',\omega)|^{2}
\end{equation}
with $\beta$ is the Bogoliubov coefficient given by
\begin{equation}
\beta(\omega' \omega)~~ =~~ {1 \over {2\pi}} \sqrt{{\omega'} \over \omega} \int_{-\infty}^{V_0}
dV e^{-i{\omega'} V} e^{-i \omega {\cal F}(V)}.
\end{equation}
Here, the function $U={\cal F}(V)$ is the inverse of the map $V={\cal G}(U)$. The ingoing ray
$V_{0}$ is the last ray which becomes an outgoing ray after passing through the center of the
star. In case of the naked singularity model, one needs to make a subtle assumption about the
nature of emission from beyond the Cauchy horizon. The present analysis amounts to the assumption
that there are no outgoing modes from beyond the Cauchy horizon \cite{Vaz}. In principle, one
could consider alternate boundary conditions on and in the future of the Cauchy horizon (see
for instance Harada et al. \cite{Har}).

It follows that,
\begin{equation}
\langle 0|N(\omega)|0\rangle=
{1 \over {4\pi^2}\omega}
\int_{-\infty}^{V_0} dV \int_{-\infty}^{V_0} dV'  \int_0^\infty d\omega'
\omega' e^{i{\omega'}(V'-V)} e^{i\omega[{\cal F}(V')-{\cal F}(V)]}
\end{equation}
and carrying out the integral over $\omega'$ we find
\begin{equation}
\langle N(\omega)\rangle=  -{1\over 4\pi^{2}\omega} \int_{-\infty}^{V_{0}} dV
\int_{-\infty}^{V_{0}}
dV'
{1\over
(V'-V + i\epsilon)^{2}}~ e^{i\omega [{\cal F}(V')-{\cal F}(V)]}.
\label{eps}
\end{equation}
Now inverting the expression (\ref{bhm}) for ${\cal G}(U)$ to obtain the map ${\cal F}(V)$
for the black hole, and making the substitutions
$x=V_{0}-V$, $y=V_{0}-V'$, and then $p=x-y$, $q=y/x$  we get
\begin{equation}
N(\omega)=-{1\over 4\pi^{2}\omega} \int_{-\infty}^{\infty} {dp\over p}
 \int_{0}^{\infty} {q^{-4iM\omega}\over (1-q + i\epsilon)^{2}}.
\end{equation}
The integral over $q$ can be performed \cite{Gra}. Finally taking the $\epsilon\rightarrow
0$ limit, we obtain
\begin{equation}
N(\omega)={2M\over \pi} {1\over e^{8\pi
M\omega}-1}\int_{-\infty}^{\infty} {dp\over p}.
\end{equation}
Thus we recover the expected thermal result for the frequency dependence in the case of the black
hole. The divergence resulting from the integral over $p$ represents a steady flux over an infinite
time and can be eliminated by considering the emission per unit time.

The integral for the naked singularity case can be performed by inverting the map given in (\ref{nsm})
and by making the substitutions $x=V_{0}-V$, $y=V_{0}-V'$ followed by $p=x^{\gamma}-y^{\gamma}$, $q=y/x$.
This gives
\begin{equation}
N(\omega)=-{1\over 4\pi^{2}\omega} \left [ \int_{0}^{\infty} dp
\int_{1}^{\infty} dq + \int_{-\infty}^{0} dp \int_{0}^{1} dq \right ]
{e^{i\omega B p}\over \gamma |p|} {1 \over (1-q+i\epsilon)^{2}}.
\end{equation}
or, after integrating,
\begin{equation}
N(\omega) \approx {\epsilon^{-1}\over 4\pi\omega\gamma}.
\end{equation}
The frequency dependence of the spectrum is $1/\omega$. In evaluating the integral (\ref{eps}) for
the naked case we have made the change $\epsilon\rightarrow\epsilon x$. As is seen, the role of
the regulator $\epsilon$ in the naked case is quite different from that in the black hole case. The
limit $\epsilon\rightarrow 0$ yields a divergence - the origin of this divergence appears to be different
from the black hole case, since now the flux is not for an infinite time.

The fact that the divergent behaviour of the spectrum does not depend on the parameter $\gamma$
characterizing the naked singularity suggests that this spectrum may be the result of imposing
self-similarity on the solution. It could also be related to the cutoff at the Cauchy horizon (i.e.
the assumption that there are no outgoing modes beyond the Cauchy horizon). It also suggests that an
inference of the spectrum will strongly depend on the nature of boundary conditions imposed on the
Cauchy horizon.

We would like to emphasize that the boundary conditions assumed here for the spectrum calculation are
not at all relevant for the calculation of the radiated energy described in the next section. The result
on the radiated energy is entirely based on calculations in the spacetime region prior to the Cauchy
horizon. Hence there is no need to make any assumptions regarding boundary conditions on the Cauchy
horizon, in so far as the calculation of emitted energy is concerned.

\section{Radiated Energy}

Let us now calculate the radiated energy. Using (\ref{pow}) and the forms (\ref{bhm}) and (\ref{nsm})
for $\cal{G}(U)$ we get the radiated power in the black hole case as
\begin{equation}
P = \hat{P}={\hbar \over 768\pi M^{2}}
\end{equation}
and in the naked case as
\begin{equation}
P = {\hbar \over 48\pi}
\left[{1- \gamma^{2}\over  \gamma^{2} (U_0-U)^{2}}\right],
\label{ouch}
\end{equation}
and
\begin{equation}
\hat{P}={\hbar \over 48\pi}
\left[\frac{(1- \gamma)^{2}}{\gamma^{2} (U_0-U)^{2}}\right].
\end{equation}
The net, integrated, flux emitted in the naked singularity model can be written as
\begin{equation}
E=\int_{-\infty}^{U} P(U') dU'  = {\hbar \over 48\pi}
\left[{1- \gamma^{2}\over  \gamma^{2} (U_0-U)}\right],
\label{ene}
\end{equation}
 and
\begin{equation}
\hat{E}=\int_{-\infty}^{U} \hat{P}(U') dU'  = {\hbar \over 48\pi}
\left[\frac{(1- \gamma)^{2}}{\gamma^{2} (U_0-U)}\right].
\end{equation}

As seen above, in the limit that $U$ approaches the Cauchy horizon $U_{0}$, the emitted flux
diverges. However, this divergence needs to be interpreted carefully. We are working with the
semiclassical approximation, in which the gravitational field has not been quantized. The
semiclassical approximation ceases to be valid if the curvature at some point inside the star
approaches Planck scale. From (\ref{den}) we see that this happens at the center of the star at
$t\approx -t_{Planck}$ where $t_{Planck}$ is the Planck time. Hence, we may say that the results
of the semiclassical approximation cannot be trusted beyond this epoch.

From equations (\ref{tee}) and (\ref{ooo}) one can show that for a ray starting from the origin
at $t=-t_{Planck}$, the time difference $U_{0}-U$ is greater than of the order of Planck time.
This is done as follows. Let ray $A$ start from the origin $r=0$ at $t=-t_{pl}$ and ray $B$
start from the origin at the singular epoch $t=0$. The outward propagation of a radial null ray
is described by the equation
\begin{equation}
{dt\over dr}=R'={ {1-{\sqrt {\lambda}\over 2}{t\over r}} \over
\left({1-{3\sqrt {\lambda}\over 2}{t\over r}}\right)^{1/3} }.
\label{dtdr}
\end{equation}

Since ray $B$ arrives at a comoving coordinate  $r$ later than ray $A$, it follows from this
equation that the value of $dt/dr$ at a given $r$ will be higher for $B$ compared to $A$. Hence
the time interval $\Delta t\equiv t_B-t_A$ between the arrival times $t_B$ and $t_A$ of these
two rays at the boundary $r_b$ of the star will exceed $t_{pl}$.

Using equation (\ref{ooo}) and the definition of $y$ we can write the difference $\Delta U
\equiv U_0-U\equiv U_B-U_A$ as
\begin{equation}
\Delta U= t_B-t_A  + {\rm other\ positive\ terms}
\label{diff}
\end{equation}
This shows that the asymptotic time difference $U_0-U$ will be greater than of the order of
Planck time. Hence, one can see from (\ref{ene}) that in the range that the semiclassical
approximation can be trusted, the emitted semiclassical flux is less than of the order of
Planck energy, which is in general much smaller than the mass of the original star.

This suggests that very early during the collapse, the star enters the quantum gravitational
phase, with most of its mass intact. The further evolution, including the nature, duration and
amount of emission, will be determined by quantum gravity. In this sense, the naked singularity
system behaves very differently from the black hole system. For an evaporating black hole, quantum
gravitational effects become important only during the very final stages, when the mass of the
black hole approaches Planck mass. For an astrophysical black hole, the time taken to reach this
quantum gravitational phase is enormous ($10^{71}(M/M_{\odot})^{3}\mbox{s}$). Also, it is an open
issue as to whether the black hole evaporates entirely, or leaves behind a Planck mass remnant.

On the other hand, for the star forming a naked singularity, quantum gravity becomes important much
before the completion of the collapse. And it is at present an open issue as to whether the naked
star settles into a black hole or explodes on a very short time scale. This issue has to be
decided by quantum gravity. Perhaps one can say that for the first time physicists have encountered
a dynamical system whose evolution, within the lifetime of the Universe, can only be understood
with the help of quantum gravity. Indeed, the Universe may well contain such stars, whose final
evolution cannot be understood without quantum gravity. A study of such systems appears to be a
useful probe of theories of quantum gravity like canonical general relativity and string theory.

\section{Back Reaction}

An important aspect of the above analysis is that we have ignored the back-reaction, and one could
well ask if our conclusions could be affected by its inclusion. We now give arguments as to why
the back reaction cannot be important during the semiclassical evolution, in the present case. One
can propose two different criteria for deciding as to when the back reaction becomes important. One
is that the total flux received at infinity becomes comparable to the mass of the collapsing star.
As we have seen above, if the mass of the collapsing star is much greater than the Planck mass (as
is of course usually the case) then the back reaction does not become important during the
semiclassical phase, because the net emission cannot be more than about one Planck mass.

The second criterion for the back-reaction to become important is that the energy density of the
quantum field becomes comparable to the background energy density, {\it inside} the star. In a
four-dimensional model, it is difficult to establish whether this happens. However, one can study
the evolution of the quantum field inside the star using the 2-d model obtained by suppressing the
angular part of the 4-d Tolman-Bondi model \cite{Bar2}.  Using this model, Iguchi and Harada
\cite{Igu} have shown that during the domain of validity of the semiclassical approximation the
expectation value of the energy momentum tensor of the scalar field remains smaller than the
energy density of the dust matter in the collapsing background. Thus the back reaction does not
become significant during the semiclassical evolution, inside the star, for this 2-d model. Since
a solution of the four dimensional model obtained using the geometric optics approximation is
equivalent to the exact solution of the two-dimensional version, it is plausible that this result
holds for the four dimensional stellar dust model as well. In this case we can conclude that quantum
gravitational effects are more important than the semiclassical back reaction in deciding the
evolution of the star.

Again we see that the semiclassical evolution is very different from the black hole case, since
the back reaction plays a crucial role during the semiclassical history of the black hole.

\section{The Smooth Case}

Similar conclusions about the nature of the evolution hold also for a $C^{\infty}$ model studied by
Harada et al. \cite{Har}. They also find that the amount of energy emitted during the semiclassical
phase is of the order of Planck energy. Hence our results are not specific to the self-similar model.

The situation is similar but more complicated than the self-similar case. The $C^{\infty}$ case is given by
\begin{equation}
F(r)=F_{3}r^{3}+F_{5}r^{5}+\cdots.
\end{equation}
Then Harada et al. found that the function $\cal G$ behaves as
\begin{equation}
{\cal G} =V_{0}-A(U_{0}-U)-f \nu_{s}^{1/2}(U_{0}-U)^{3/2}+\mbox{higher order terms},
\end{equation}
where $\nu_{s}$ is determined by the initial density profile and the above formula is valid only for
$\nu_{s}(U_{0}-U)<1$.
Using the above formula, they also found
\begin{eqnarray}
P&\sim& \hbar \nu_{s}^{1/2}(U_{0}-U)^{-3/2}, \\
\hat{P}&\sim& \hbar \nu_{s}(U_{0}-U)^{-1}, \\
E&\sim& \hbar \nu_{s}(U_{0}-U)^{-1/2}, \\
\hat{E}&\sim &\hbar\nu_{s}\ln [\nu_{s}(U_{0}-U)^{-1}],\\
N(\nu)&\sim& \nu_{s} \nu^{-2}.
\end{eqnarray}
It is noted that $\nu_{s}$ may exceed the Planck frequency $\nu_{Planck}$ because $\nu_{s}$ is determined
by some combination of two macroscopic scales, the free-fall time and the scale of inhomogeneity.
Nevertheless, we can find that the total radiated energy for this late time radiation cannot exceed the
Planck energy $\hbar \nu_{Planck}\sim 10^{16}\mbox{erg}$ before the Planck time $t_{Planck}=10^{-43}
\mbox{s}$. The proof is the following. If $\nu_{s}>\nu_{Planck}$, then it is found that the late time
radiation is only for $U_{0}-U<\nu_{s}^{-1}<t_{Planck}$. Therefore, we only consider $\nu_{s}<
\nu_{Planck}$. For this case, it is found
\begin{eqnarray}
E_{U_{0}-U\sim t_{Planck}}&\sim& \hbar (\nu_{s}\nu_{Planck})^{1/2}<\hbar \nu_{Planck}, \\
\hat{E}_{U_{0}-U\sim t_{Planck}}&\sim &\hbar\nu_{Planck}
\left(\frac{\nu_{s}}{\nu_{Planck}}\ln \frac{\nu_{Planck}}{\nu_{s}}\right)
<\hbar \nu_{Planck},
\end{eqnarray}
where the second inequality comes from the fact that the function $-x \ln x$ has a maximum value $e^{-1}$
in the domain $0<x<1$.

\section{Discussion}

One could conjecture that this picture (i.e. the semiclassical flux is of the order of Planck energy) will
in general hold for a naked singularity forming in spherical collapse, even if the matter model is other
than dust. This is because the curvature induced quantum particle production is a kinematical effect and
the only role dust plays is that of providing the background gravitational field. Any other matter model
giving rise to a spherical naked singularity can be expected to have a similar semiclassical evolution. The
fact that the semiclassical emission is so small appears to suggest that the emission comes from a very
small central spacetime region in the neighborhood of the naked singularity. If so, then this is a property
of the spherical geometry, which should continue to hold when pressure is included. In the case of a
black hole, the semiclassical emission is a property of the event horizon - as a result the evaporation
rate is slower and the semiclassical phase lasts for a much longer duration.

In contrast with the spherical case, if the collapse is highly non-spherical, say resulting in a spindle
naked singularity, it is possible that the semiclassical emission could be a significant fraction of the
initial mass. This is an important open problem which deserves to be investigated.

Understanding the quantum gravitational evolution of a spherical star which terminates in a naked
singularity is thus an outstanding challenge for candidate theories of quantum gravity. One would like to
know whether the star settles into a black hole state or instead, undergoes a catastrophic explosion on
a short time scale, resulting in a cosmic burst. A first step towards modelling this phase has been taken
in \cite{vazmidi} where a formalism is developed for the midisuperspace quantisation of the Wheeler-deWitt
equation describing spherical dust collapse.
\vskip 3mm

\noindent{\bf Acknowledgements}

\noindent TH is supported by the Grant-in-Aid for Scientific
Research (No. 05540) from the Japanese Ministry of
Education, Science, Sports and Culture.
KN is supported by Grant-in-Aid for
Creative Basic Research (No. 09NP0801) from Japanese Ministry of Education, Science, Sports and Culture.
TPS and CV acknowledge the partial support of the FCT, Portugal, under contract number POCTI/32694/FIS/2000.
TT is supported by the Monbusho Grant-in-Aid No. 1270154.

\end{document}